\documentclass[11pt,conference]{IEEEtran}
\usepackage[utf8]{inputenc}
\usepackage{graphicx}
\usepackage{balance}
\usepackage{xcolor}
\usepackage{amsmath}
\usepackage{footnote}
\usepackage{url}
\usepackage{hyperref}
\usepackage{tabularx,ragged2e,booktabs,caption}
\newcolumntype{C}{>{\Centering\arraybackslash}X}
\makesavenoteenv{tabular}
\usepackage[draft]{todonotes}   

\title{Bioinformatics Computational Cluster Batch Task Profiling with Machine Learning for Failure Prediction}

\author{\IEEEauthorblockN{Christopher Harrison\IEEEauthorrefmark{1}\IEEEauthorrefmark{2} Christine R. Kirkpatrick\IEEEauthorrefmark{3}\IEEEauthorrefmark{2}
and In\^es Dutra\IEEEauthorrefmark{2}\\}

\IEEEauthorblockA{\IEEEauthorrefmark{1}
Department of Biostatistics and Medical Informatics\\
School of Medicine and Public Health\\
University of Wisconsin - Madison\\
Madison, Wisconsin USA
}
\IEEEauthorblockA{\IEEEauthorrefmark{2}
Departamento de Ci\^{e}ncia de Computadores\\
Faculdade de Ci\^{e}ncias\\
Universidade do Porto\\
Porto, Portugal
}
\IEEEauthorblockA{\IEEEauthorrefmark{3}
San Diego Supercomputer Center\\
University of California San Diego\\
La Jolla, California USA
}

}
\date{October 2018}

\begin{document}

\maketitle

\begin{abstract}
Motivation: Traditional computational cluster schedulers are based on user inputs and run time needs request for memory and CPU, not IO. Heavily IO bound task run times, like ones seen in many big data and bioinformatics problems, are dependent on the IO subsystems scheduling and are problematic for cluster resource scheduling.  The  problematic rescheduling of IO intensive and errant tasks is a lost resource. Understanding the conditions in both successful and failed tasks and differentiating them could provide knowledge to enhancing cluster scheduling and intelligent resource optimization.  

Results: We analyze a production computational cluster contributing 6.7 thousand CPU hours to research over two years.   Through this analysis we develop a machine learning task profiling agent for clusters that attempts to predict failures between identically provision requested tasks.

\end{abstract}
\section{Introduction}
Resource allocation is a well studied problem. Solutions have been sought for different kinds of applications and platforms, ranging from multi-core centralized memory machines to distributed memory machines, passing by many-core and GPGPUs. The objective is always the
same: to better allocate resources (CPU, memory and storage) in order
to improve some performance metrics (response time, throughput,
speedup, efficiency, Quality of Service (QoS), among others). Due to specific
characteristics of applications and platforms, it is not possible to
devise one solution that fits all. It is also not possible to devise
solutions that simultaneously will improve all metrics. With the ever
growing use of cloud computing platforms and their virtual machines,
new solutions have been studied. Monitoring tools have been built that
can help notify system administrators to address jobs which need to be rescheduled, restart, create, or destroy and these tools extend to 
virtual machines as a way to also keep these systems running. These tools also help operators justify the need for additional compute resources.

Cloud computing platforms rely on clusters of machines to run users
jobs and their middleware provides an interface (command line, API or
portal) to help submitting single or batch jobs. Usually, users need
to specify their job's requirements such as expected execution time,
expected memory usage, storage, number of cores, among
others. However, the fact is that since the creation of the first
parallel and distributed batch platforms, most submitted jobs fail to
complete for several different reasons. The high rate of failure is an
impediment to the good use of resources. Statistical
analysis of trace logs were performed by Prodan and
Fahringer~\cite{Radu}, and by Christodoulopoulos {\it et al.}, among
others, where conventional algorithmic solutions have been used to
improve resource allocation and prevent failures. More sophisticated
solutions, based on machine learning techniques, have also been used
to identify patterns of success and failure and to optimize resource
usage~\cite{4579676}. We study traces of a two-year
period of jobs submitted using the Condor HTC tool with the objective
of understanding patterns of success and failure. These jobs were all
submitted by biostatisticians or medical informatics groups. We
analyzed all jobs and identified two kinds of submissions: jobs with a
single task submission and jobs with multiple tasks submissions. While
single task job submissions are very common, they have very specific
characteristics that make it difficult to find patterns. Therefore, we
concentrated on the multiple tasks submissions jobs, since they tend
to share common characteristics such as the same command and possibly
the same requirements. 
We categorized these jobs in five mainstream groups: 
\begin{enumerate}
    \item  jobs submissions of single tasks which finished successfully
    \item  jobs submissions of single tasks which failed
    \item  jobs with multiple task submission for which all tasks finished successfully
    \item jobs with multiple task submission for which all tasks failed
    \item jobs with multiple task submission for which some of the tasks
succeeded and some failed
\end{enumerate}
 With the last group, jobs with multiple task submission for which some of the tasks
succeeded and some failed, we applied a machine
learning technique to each job in order to identify patterns of
success and failure among their tasks. The average number of tasks per
job of this kind is 182.3. Results show that an average of 14.3\% fail and
85.7\% succeed. The jobs that fail all have in common a HTCondor class ad `JobStatus=3` whereas Jobs that succeed all have in common a HTCondor class ad `JobStatus=4`. By detecting such patterns, we expect to use them to detect future new tasks likely to fail that
enter the system, and adjust or correct them in order to minimize
failures and optimize resource usage.

\section{Background}
Computational Cluster task scheduling, such as ones used in HPC/HTC systems, distribute tasks based on available HPC/HTC resources. The task schedulers match user requested resources per task to the expected resource available.   If a task resource request is not sufficient to complete the task then additional task resource allocations are made based on the scheduled cluster host resource.  The goal is to efficiently allocate resources based on task matched resource requests and optimize based on requested demand.   As such, an HPC/HTC scheduler has two optimization areas which are: 
\begin{enumerate}
    \item granular measurements of process needs: IO, CPU, network, system wait times, etc.
    \item  enhancing the end user's task estimating needs based on a best guess.
\end{enumerate}

System and task monitoring of performance metrics are essential for evaluating resource allocation and utilization.   These long term trends based on system analysis compared to base lines are widely used as a justification for additional resources.  However, optimization of resources is difficult without granular resource task trends. Resource usage statistics and trends tools are designed to lose fidelity over time due to the amounts of data generated by performance metrics.   Additionally, the amount of data collected for metrics grows as a function to cluster size.    While HPC resource monitoring platforms exist; such as ganglia~\cite{gangliamon}, the value of their longitudinal data decreases over time due to the application storage formats of their metrics.  Specifically RRD’s (Round-Robin Databases) use results in degrading signal value over time.

We monitor cluster resource utilization with granular metrics for scheduling and running tasks.
Our problem is to optimize resources utilization through modeling of failed tasks through monitoring and adaptively responding to task resource needs based on available cluster resources.   Similarly to Juve {\it et al.}~\cite{7307664}, we profile our cluster tasks while addressing resource loss as a function of time and impact to total resource utilization.   Specifically, we focus on cluster resource loss as a function of failed tasks which have identical attributes to other successful tasks, also known as a same cluster submission job.   Our problem is complicated due to temporal computational cluster resources utilization needs which fluctuates based on scheduled running tasks, service outages, unplanned outages, network anomalies, etc.   All submitted tasks are impacted by Spatio-temporal phenomena\cite{1751-8121-50-10-103001} with time impacting usage, and queued demand impacting temporal run time and scheduling.

Understanding and tracking performance metrics over time with consistent time series granularity is resource intensive.   At scale, systems designed for simple collection and analysis of groups of systems cannot handle the influx of 10 to 100 of thousands records a second while being real time search-able.   Inefficient resource utilization equals wasted resources; therefore, gauging improvements to application and system performance will provide a baseline for resource optimization. Cluster tasks requiring a reschedule are wasted resources due to the cost of compute time on a cluster.   Thus a need for better failure prediction and cluster rescheduling after task failure provides a pathway towards resource optimization.

Human best guess estimates can often be wrong as can be seen in baseball~\cite{houser2005baseball}.   Individual users can misinterpret the resources required for a given task and request more or less resources which leads to a misallocation of cluster resources and non optimal cluster utilization outcomes.

\section{Related Work}
Task profiling and system performance metrics are essential for effectively evaluating task allocations and cluster resource utilization.   Long term trends and individual analysis compared to base lines are widely used as a justification for additional resources.   When evaluating HPC system utilization a baseline analysis and long term trends are needed to justify additional resource allocations.   However, optimization of resources is difficult without granular resource trends.   Resource usage statistics and trends lose fidelity over time due to the amounts of data generated by performance metrics and grows as a function to cluster size.    While HPC resource monitoring platforms exist such as ganglia~\cite{gangliamon}, the value of their longitudinal data decreases over time due to the application storage formats of their metrics.  Specifically RRD’s (Round-Robin Databases)~\cite{rrdcache} use results in degrading signal value over time.

Numerous tools, such as cacti~\cite{cactimon}, ganglia~\cite{gangliamon}, nagios~\cite{nagiosmon}, and zabbix~\cite{zabbixmon}, exist to evaluate system utilization and track performance metrics.   These tools attempt to prove graphical representations of system performance metrics and utilize the network standard Simple Network Monitoring Protocol (SNMP)~\cite{rfc1157}.   Tracking metrics can address application bottlenecks such as application starvation or usage of memory, CPU, or IO by critically understanding resource allocation starvation and optimization at starvation time.

 OVIS~\cite{1639698} has attempted to address scheduling of resources based on predictive failure analysis but not on the user requested resource allocation.   Additionally, OVIS's scope was limited to resource allocation improvements to address task scheduling around node failures not optimizing cluster resources.    While a given node’s failure will impact scheduling to that node and is impacted by the Mean Time To Failure (MTTF), the resource scheduling to a given node is a function of the given node’s failure probability.   Thus, OVIS attempts to address the issue of resource failures within a cluster by working around hardware failures at the node level.

Larger scale system monitoring of HPC and the Open Science Grid~\cite{1742-6596-78-1-012057} have used such systems as OVIS~\cite{1639698} or TACC Stats~\cite{7081222}. OVIS uses a Bayesian inference scheme to dynamically infer models for the normal behavior of a system and to determine bounds on the probability of values evinced in the system.   OVIS addresses hardware related failure issues and system level performance analysis on systems based on MTTF analysis of a given system.

\begin{figure}[!t] 
  \centering
  \includegraphics[width=3.5in]{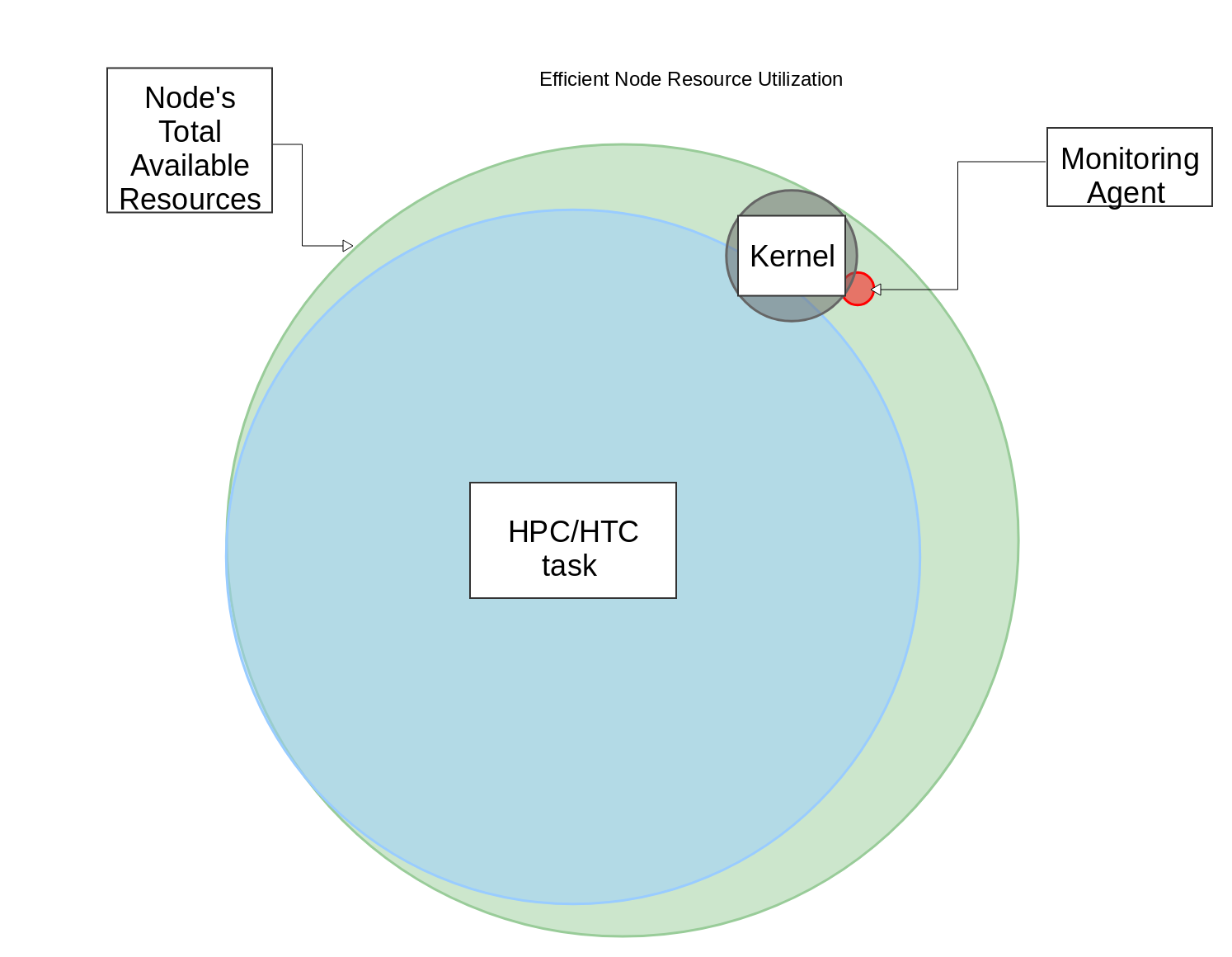}
  \caption{Efficient task run state on cluster node}
  \label{fig:efficienttask}
\end{figure}

Understanding system resource usage across a cluster, based on a per scheduled task basis instead of an entire system, requires an understanding of task profiling.   Given tasks profiles for CPU, memory, and disk have been demonstrated~\cite{Seneviratne:2011:TPM:1890010.1890174} as a method for ``choosing the most suitable set of computers for the deployment of the tasks''~\cite{Seneviratne:2011:TPM:1890010.1890174}. As such, computing resource usage measurements are required to effectively and efficiently utilize computing resources, Fig.~\ref{fig:efficienttask}.

Modeling resource utilization as a means for optimization predictions is needed in both HPC and cloud (as HTC) contexts. Even newest of breed techniques, such as the use of Kubernetes for cloud orchestration, only consider CPU in scheduling decisions. This is too simplistic for true optimization.  Newer models and software implementations are needed to more effectively understand usage and suggest future scheduling of resources.  Chang {\it et al.} suggested using multiple variables, such as memory and disk access, and creates a dynamic algorithm that can adjust to tasks in situ~\cite{8254046}. Wei {\it et al.} demonstrated a technique for allocating Virtual Machine (VM) resources based on CPU and memory~\cite{7277025}, albeit limited. Along with various approaches, systems and models schedule using different parameters and thereby optimize usage based on various priorities, such as deadline, cost minimization or maximization, such as for cloud providers, to adhere to Service Level Agreements, and to reduce power consumption~\cite{Bhavani}.  Additional work in this area has been motivated by optimization based on energy savings, where Pinheiro notes that it is key to examine resource reconfiguration and keeping a load stable (relatively unchanged) as the throughput loss can be resource intensive~\cite{Pinheiro01loadbalancing}. Because of the heterogeneous hardware and unpredictable loads in shared environments~\cite{Homer:2014:CDP:2636530}, cloud systems perhaps, need predictive scheduling and optimization more than traditional, homogeneous HPC clusters. To disregard optimization in cloud systems is to ignore the promise of the cloud as an elastic resource, made to adapt to computing loads in ways previously limited by dedicated hardware~\cite{Bhavani}. This work can be extended to heterogeneous HPC systems, as can occur in condo storage models, where nodes are purchased over time~\cite{TSCC}, as well as where users are able to bring their own nodes to the cluster. Mateescu describes a technique where scheduling is done in part based on the timing demands of the tasks and can utilize combinations of HPC and Cloud as a combined workflow, either managing at the node level (physical machine) or at the VM level~\cite{MATEESCU2011440}.

Recently, Rodrigo {\it et al.}~\cite{Rodrigo:2017:EWS:3078597.3078604} proposed a workflow aware scheduling algorithm specific to slurm\~cite{yoo2003slurm} which attempts to address temporal and locality resource scheduling.   This approach is interesting as it attempts to match pipelines, beyond individual tasks, to resources for data locality and increased throughput.   However, the resource locality effect is muted for clusters that do not have local resources (e.g., nodes without local disk for scratch space), which affect scheduled tasks in a homogeneous HPC/HTC environment.  Considering previous work in scheduling and optimization~\cite{Bhavani}, what is unique about this method is the use of machine learning to suggest optimization for real-time workloads, that works across various size systems, under a wide range of applications.  

\section{Methods}
Following procedures originally used by Juve {\it et al.}~\cite{7307664} we collect task statistics and compare our statistics against the task's requested cluster resources.  Using Juve {\it et al.} method we use the metric data polled directly from the kernel through the procfs file system. We also note that  only a subset of the proc file system, of which the kernel provides a non-intrusive task metric to gain individual process statistics about performance and run time state in real time, is used. A list of available procfs metric's is defined in~\ref{tab:procfs_attributes}. The defined list is meant to be a demonstration to the richness to which the kernel procfs  application level statistics can be distilled.\\
\begin{table}[]
    \centering
    \begin{tabular}{l|l|p{4cm}}
    \hline
    \textbf{file} & \textbf{type}  & \textbf{description} \\
    \hline
    \hline
     children & list numbers & Child tasks process id (pid) from the this task id\\
     \hline
     cmdline & text & full command line used to execute this task with options\\

     \hline
     cwd & simlink & symbolic link to the current working directory of the process\\
     \hline
     environ & text &initial environment that was set when the currently executing program was started via execve\\
     \hline
     fd & directory & a subdirectory containing one entry for each file
              which the process has open, named by its file descriptor, and
              which is a symbolic link to the actual file\\
    \hline
    fdinfo & directory &  a subdirectory containing one entry for each file
              which the process has open, named by its file descriptor\\
    \hline
    io & text &  contains I/O statistics for the process

               \begin{itemize}
                   \item rchar: characters read
                   \item wchar: characters written
                   \item syscr: read syscalls
                   \item read\_bytes: bytes read
                   \item write\_bytes: bytes written
                   \item cancelled\_write\_bytes:
                    The big inaccuracy here is truncate.
                   \end{itemize} \\
     \hline
    mounts & text & lists all the filesystems currently mounted in the
              process's mount namespace\\
    \hline
    oom\_score & int & displays the current score that the kernel gives to
              this process for the purpose of selecting a process for the
              OOM-killer\\
    \hline
    pagemap & text & the mapping of each of the process's virtual
              pages into physical page frames or swap area\\
    \hline
    smaps & text & shows memory consumption for each of the process`'s
              mappings\\
    \hline
    stat & text & Status information about the process\\
    \hline
    statm & text &  Provides information about memory usage, measured in pages \\
    \hline
    status & text & Provides much of the information in /proc/[pid]/stat and
              /proc/[pid]/statm in a format that's easier for humans to
              parse\\
    \hline
    syscall & text & exposes the system call number and argument regis‐
              ters for the system call currently being executed by the
              process, followed by the values of the stack pointer and pro‐
              gram counter registers\\
    \hline
    wchan & text & symbolic name corresponding to the location in the kernel
              where the process is sleeping\\
\end{tabular}

    \caption{procfs task files~\cite{linuxmanprocfs}}
    \label{tab:procfs_attributes}
\end{table}
Procfs provides kernel counters on a per task basis. However, these task based kernel counters have grown organically over the years and have resulted in a less than ideal task counter framework. Specifically, the procfs counter files within the procfs file system are a mixture of file formats and none are consistent.

\subsection{Computational cluster}

\subsubsection{University of Wisconsin - Madison, Biostatistics Computing Group}
The cluster is a departmental level cluster comprised of 240 64 bit multicore x86 systems running a Redhat Package Management (RPM)~\cite{RPM} based Linux variant, Scientific Linux 7~\cite{SL}.   The cluster is a heterogeneous mixture of hardware vendors and CPU brands with approximately 2500 CPU cores and 11.8 TB of total memory across the cluster. Hardware manufacturers used in the cluster include: Dell, Cisco, HPC and Supermicro with x86 CPUs.

The cluster is comprised of shared use interactive machines and dedicated compute machines. Both types of machines are part of a cluster which uses HTCondor~\cite{gridbook-htc} as the cluster resource manager.

Profiling statistics from a system are generated through the kernel. The Linux kernel procfs contains task level runtime parameters which live reflect the task runtime within the OS scheduler.  Additionally, any task’s sub task or children are also defined in the procfs, as the Parent Process Id (PPID). Any subtask task which is created by fork the system will register the child process Id (pid) under the parent as defined in the procfs path \textit{$/proc/\$ppid/task/\$ppid/children$}.   The procfs we describe is used on the cluster installed Linux variant Scientific Linux 7~\cite{SL} running a Linux kernel 3.10.x.

  The focus of the tasks we used in our analysis are all in support of the Department of Biostatistics and Medical Informatics at the University of Wisconsin - Madison.   These tasks where submitted between epoch times (1475644627 - 1537752190) or between October 5, 2016 to September 24, 2018. All submitted tasks are used by biostatisticians or medical informaticians and require compliance or have dataset restrictions which inhibit them from running on across infrastructures such as the OSG\cite{1742-6596-78-1-012057} or other open resources.   Using the HTCondor class-Ad history we analyze 17,282 cluster submissions producing 129,854 tasks in the BCG HTCondor\cite{gridbook-htc} computational cluster. Table~\ref{tab:task_submitted} shows a complete breakdown of the total tasks by completion state.   Additionally, in Table~\ref{tab:task_breakdown}, we can see the full exit status of tasks which did not complete successfully.

  \begin{figure}[!t] 
 \centering
  \includegraphics[width=3.5in]{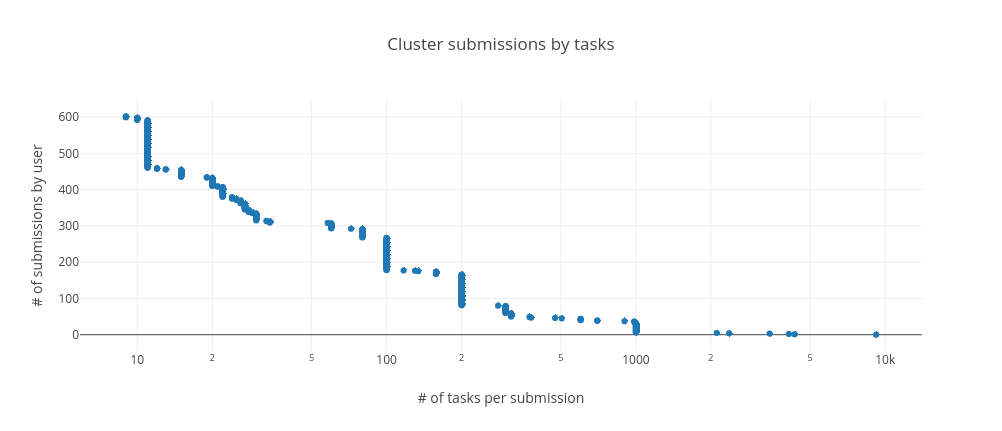}
  \caption{Number for tasks per  cluster submission}
  \label{fig:clustertask-fig}
\end{figure}

\begin{table}[]
    \centering
    \begin{tabular}{r|r|p{4cm}}
    \hline
    \textbf{number} & \textbf{percent}  & \textbf{description} \\
    \hline
    \hline
    106695 & 82.2\% & Complete successful, no errors\\
    23159 & 17.8\% & Task Error Issues\\
    \hline
    129854 & 100\% & All submitted tasks\\

  \end{tabular}
    \caption{Submitted Task Breakdown}
    \label{tab:task_submitted}
\end{table}

\begin{table}[]
    \centering
    \begin{tabular}{r|r|p{4cm}}
    \hline
     \textbf{number} & \textbf{percent}  & \textbf{event}\\
     \hline
    \hline
    13070 & 10.1\% & User removed before scheduled to cluster\\
    9183 & 7.1\% &  User defined task attribute expression error\\
    434 & 0.33\%& Failed to initialize user log\\
    229 & 0.17\% & Out-of-memory event\\
    160 & 0.11\% & Other \\
    83 & 0.06\% & No such file or directory\\
    \hline
    23159 & 17.8\% & Total Task Error Issues\\

  \end{tabular}
    \caption{Submitted Tasks with errors breakdown}
    \label{tab:task_breakdown}
\end{table}

\begin{table*}[]
    \centering
    \begin{tabular}{r|r|r|r|l|p{4cm}}
    \hline
     \textbf{\# submission} & \textbf{\% submission} & \textbf{\# tasks}  & \textbf{\% total tasks} & \textbf{submission type} & \textbf{description}\\
     \hline
    \hline
    10758 & 62.2\% & 10758 & 8.3\% & single & successfully completed\\
    4285 & 24.8\% & 4285 & 3.3\% & single & failed\\
    1860 & 10.8\% & 93908 & 72.3\% & multi &  combo success and failed \\
    341 & 2.0\% & 16356 & 12.6\% & multi & all successfully completed\\
    38 & 0.2\% & 4547 & 3.5\% & multi & all failed\\
    \hline
    17282 & 100\% & 129854 & 100\% & & total cluster submissions\\

  \end{tabular}
    \caption{Cluster submissions breakdown}
    \label{tab:clusterid_breakdown}
\end{table*}


\subsubsection{Data exploration}
We produced our data as an aggregate from all cluster nodes which are allowed to submit to the cluster.   The data was generated through the use of condor\_history with the --json option which produced the Condor history data in JSON format.   Using the preformatted JSON HTCondor history data partitioned by individual class ad and generating a unique file for each Condor submission node, we aggregate the data by importing the data into a NoSQL database.   MongoDB~\cite{mongo} was chosen as the data store for exploring and analyzing the data due to the simplistic import nature and well established user base with a feature rich NoSQL language.

To analyze the data we used the MongoDB internal language and were able to aggregate data clusters using the condor class-Add 'ClusterId' as the unique cluster key. The cluster task resource breakdown by task and totals are defined in~\ref{tab:clustersubmission_breakdown}.  For our final dataset, we apply a split frame approach to randomly assign our data to different buckets for use in the applied machine learning. Specifically, we split frame our data based on the ratios 60\%, 30\%, 10\%.

\subsubsection{Feature selection}
To understand why our tasks either succeeded or failed by submission we applied random forest~\cite{louppe2014understanding}, a well known machine learning technique, as a classifier estimator.   We utilize the machine learning platform H2O~\cite{cook2016practical} to build our models. In choosing  the Random Forest algorithm, we address the classification and regression predictions by using the average prediction trees to make the final prediction based on class or numerical values. Thus, we address our multi-modal multi-class dataset without excessive normalization's so we can integrate: numerical, binomial, and categorical data.

\section{Experiment}

  Our cluster had a total of 15,043 unique cluster submissions where only one task was submitted to the cluster by the job submission. A breakdown of cluster submission by task result(s) is found in~\ref{tab:clusterid_breakdown} and we plot our data based on tasks per submission. Of the remaining 1,860 cluster submissions, 491 of these have greater then 5 tasks submitted per cluster submission and have both failed and successful completed tasks. We set the cutoff at 5 tasks per submission so we have enough data points to apply a machine learning algorithm, random forest, for feature selection.    We focus this analysis on the submissions of 5 or more tasks with a combination of successful and failed tasks for multi-tasked submissions, the total number of tasks submitted with 5 or greater is a subset from the last line in~\ref{tab:clusterid_breakdown}.   The total number of remaining tasks is 90,062 of the 93,908 or 69.4\% of the total cluster tasks.   We use 90,062 as our base data from which to train, validate, and test.

  \begin{table*}[]
      \centering
      \begin{tabular}{r|l|l|l|l|p{4cm}}
      \hline
       \textbf{Usage statistic} & \textbf{Cumulative total} & \textbf{Average} & \textbf{Max} & \textbf{Min} & \textbf{Description}\\
       \hline
       \hline
        CPU  & 3522219511 & 27124.45 & 43112204 & 0  & Cumulative CPU time per task, min\\
        System CPU & 157853822 & 1358.85 & 366756 & 0  & Cumulative CPU time in system/kernel time, min \\
        User CPU & 2415838569 & 20796.25 & 19679010 & 0 & Cumulative CPU time in User space time, min \\
        Suspension  & 146800 & 1.13 & 12895 &  0 & Cumulative time a task is suspended/held, min \\
        BytesSent & 36400253248 & 295044.68 & 1475104768 & 0 & \\
      \hline
        \multicolumn{6}{p{.8\textwidth}}{\*Note: System and User CPU time are a subset of the total due to a HTCondor version change over the course of the collected dataset.}
        \end{tabular}

          \caption{Cluster task usage in minutes}
          \label{tab:clustersubmission_breakdown}
      \end{table*}

  \begin{figure}[!t] 
 \centering
  \includegraphics[width=3.5in]{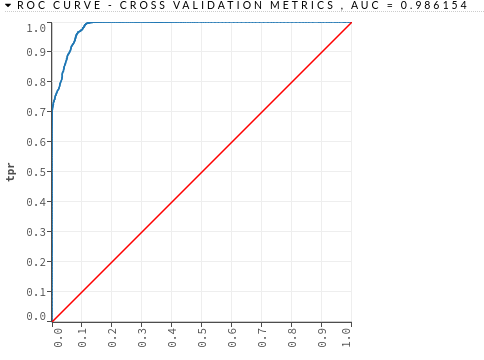}
  \caption{ROC curve, cross validation}
  \label{fig:roccrossval}
\end{figure}

  \begin{figure}[!t] 
 \centering
  \includegraphics[width=3.5in]{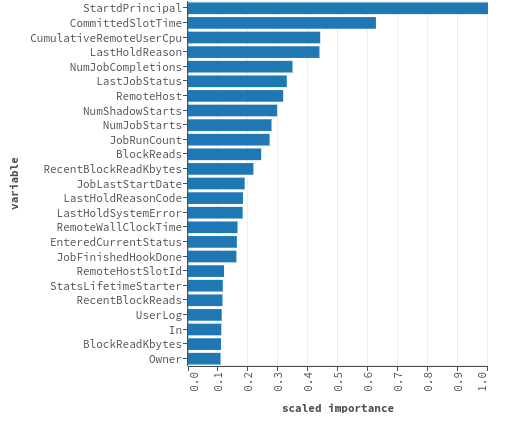}
  \caption{Variable Importance}
  \label{fig:varimport}
\end{figure}

\subsection{Prediction}
To build our Random Forest we select 50 trees, each with a max depth of 50.    Number of variables randomly sampled as candidates at each split was set at 5 ({\bf mtries}) from the active predictors requests within H2o.ai~\cite{H2o}.  Additionally, we chose to build histograms for numerical columns and chose to use histogram bins at 1000.   Addressing cross validation of our random forest trees we chose a cross validate fold number of 5, with a random fold assignment.   The number of categorical and top level histogram bins was set at 1024.   For our distribution function we utilized a multinomial distribution when building our model.   Lastly, we ignored columns when building our model, namely the columns: \textit{id}, \textit{AutoClusterId}, \textit{CommittedTime}, \textit{CompletionDate}, \textit{ExitCode}, \textit{LastVacateTime}, \textit{RemoteSysCpu}, \textit{RemoteUserCpu}, \textit{RemoveReason}.   On our initial feature selections tests, these columns confounded feature selection which heavily skew our initial results.

Using our random forest model, we were able to obtain a 99\% precision for failed tasks and a 94\% precision for successful tasks. Of these tasks we had an error rate of 11.6\% for failed tasks and a 0.5\% error rate for successful tasks.   For recall during our cross validation we saw a 88\% recall of failed tasks and 1.0 for successful tasks.   The total error rate was 4.6\%.    We were able to predict the relative importance of each attribute within the HTCondor class-Ad with scaled importance. Deterministically, our class-Ad attributes for predicting failure vs. success is ranked by contribution according to~\ref{fig:varimport}: \textit{StartdPrincipal} (12.1\%), \textit{CommittedSlotTime} (7.6\%), \textit{CumulativeRemoteUserCPU} (5.3\%), \textit{LastHoldReason} (5.3\%), \textit{NumJobCompletions} (4.2\%), \textit{LastJobStatus} (4\%), \textit{RemoteHost} (3.8\%), \textit{NumShadowStarts} (3.6\%), \textit{NumJobStarts} (3.4\%), \textit{JobRunCount} (3.3\%), \textit{BlockReads} (2.9\%), \textit{RecentBlockReadKbytes} (2.6\%), \textit{JobLastStartDate} (2.3\%).   
\textit{CommittedSlotTime} (second most valuable feature) most likely would be zero for tasks which where scheduled to a cluster node but where unable to access the users data (network or file system issue on a host).  Thus, our proposed model may still have internal feature selection confounding issues.

\section{Conclusions}
Computational cluster analysis and failure prediction is based on temporal events with identically provision requested tasks is a significant concern.   72\% of all muli-task clusters submissions had at least one task fail on our production cluster.   Understanding why identically provision requested tasks succeed or fail on a heterogeneous cluster has real world benefits for both our cluster resource scheduling and to the end users. We address the failure question through a machine learning feature selection process.  The feature selection is based on the random forest algorithm and is unique due to the significant number of production task submission we use as data for our machine learning inputs.   Our results showcase an overall accurate model ({\bf 95.4\%}) but lagged in prediction for failed jobs at 88\%.  Our use of a feature selection method provides a baseline for other production computational cluster analysis.   As with all real world example cases, normal production clusters idiosyncrasies likely impacted the specific results of our model.   

\section{Future Work}
This method could be further refined and extended through the analysis of other clusters and job profiles and address job/task confounding.  In particular, we plan to apply this technique to the analysis of other private cloud (OpenStack), to inform better service delivery and optimization of resources.  With these resource recommendations, further work should be done to implement their use to combine cloud (HTC) and dynamic HPC scheduling slots. Additionally, the linux procfs is a rich source of task resource usage data which should be explored further for resource usage scheduling optimization.  Lastly, further refinements to the feature selection process and combining it with other temporal data should result in a `job cluster effect` which will provide a more generalizable model for cluster task scheduling.

\balance

\bibliographystyle{IEEEtran}
\bibliography{bibtex.bib}
\end{document}